\documentclass[english]{cccconf}

\usepackage{amsmath}	
\usepackage{graphics}
\usepackage[comma,numbers,square,sort&compress]{natbib}
\usepackage{epstopdf}
\usepackage{color}
\usepackage{multirow}
\usepackage{array}
\usepackage{subfigure}
\usepackage{float}

\begin{document}

\bibliographystyle{unsrt}

\title{Analysis Method of Strapdown Inertial Navigation Error Distribution Based on Covariance Matrix Decomposition}

\author{Xiaokang Yang,
        Gongmin Yan,
        Fan Liu,
        Bofan Guan,
        Sihai Li}

\affiliation{School of Automation, Northwestern Polytechnical University, Xi'an, China. \email{yangongmin@163.com}}

\maketitle

\begin{abstract}
	Error distribution analysis is an important assistant technology for the research of SINS(Strapdown Inertial Navigation System). Error distribution result can provide the contribution of different errors to final navigation error, which is helpful for modifying and optimizing SINS. To realize decomposing the navigation error into parts that caused by each error source, the SINS error state space model is established and covariance matrix is decomposed according to error sources. The proposed error distribution analysis method based on 34-dimension SINS error model can quantitatively analyze the contribution to the end navigation error of initial errors, IMU(Inertial Measurement Unit) bias, IMU scale factor errors,  mounting errors of gyroscopes and accelerometers, and IMU stochastic errors. The simulations in static condition and single axis rotation condition indict that the distribution result of proposed analysis method accords with the law of error propagation. After trajectory determined, the corresponding error distribution result will be calculated with the proposed method. Compared with  the Monte-Carlo method and other method based on covariance matrix, the proposed method uses more complete error model, considers the interaction effect of error sources and can be easily realized with less computation.
\end{abstract}


\section{Introduction}
Error distribution is an important technology to analysis the influence of different error source to the final navigation error. After the vehicle's trajectory determined, the navigation error distribution, which can be used to guide the direction of IMU error restraint and optimize the SINS algorithm in purpose, of error sources will be worked out. The error distribution result is a one-dimensional table that contains proportion of each error source in final navigation result. Restraining the primary error source will effectively improve the navigation performance. Hence, error distribution technology of SINS is significant for the study on improving SINS accuracy.

Improving the accuracy of SINS is always one of important tasks for the study of navigation and guidance. Mainstream technology contains IMU error modeling, IMU precise calibration, revolution-modulation technology, optimizing SINS navigation algorithm etc\cite{Guoqiang2009,Du2016,Poddar2017,Srinivas2018}. For these methods, the primary error source of SINS should be determined firstly to assistant finding the best modified direction. It is difficult to improve accuracy of SINS without error distribution, because the complex SINS error model with more than 30 error sources need to be analysis and find out proper way to restrain them.

At present, the error distribution method can be divided into three kinds. The first is calculating navigation error with determined error model and trajectory data. This method is proper for short time situation with a simplified SINS error model. Zhang\cite{Zhang2011} deduced a tactical IMU error model and calculated the influence to position error from each error source. However, this method doesn't consider the coupling relationship of different errors. Petritoli\cite{Petritoli2021} also study on how the error evolves over time in case of simplification of calculation. The second is Monte-Carlo simulation method, which obtain the statistic result of error distribution after lots of simulation tests. Kwon\cite{Kwon2005} compared the simulation of common error model and the simulation of error model with gravity error to measure the influence of gravity error. Repeat tests must be finished because of the IMU stochastic error. If other errors need be analyzed, groups of Monte-Carlo simulation tests must be completed. The third is covariance matrix analysis, which use the covariance matrix of Kalman filter to obtain the error variance and distribution. Savege\cite{Savage2000} proposed the error distribution method for integrated navigation system by modifying the updating equation of the covariance matrix. However, the corresponding error distribution analysis of high-accuracy SINS has not been seen in the literature. Weng\cite{Weng2015} proposed an error distribution method of SINS and verified in the in-flight alignment condition. Yan\cite{Yan2018} deduced another error distribution analysis based on covariance matrix according to the updating equation of covariance matrix of Kalman filter. It can be applied in both of pure inertial navigation system and integrated navigation system. Although the method is simple and feasible, the deduction process is so complex and tedious.

A simplified and feasible error distribution analysis technology, that is based on 34-dimension SINS error model and its linear state space model, is proposed herein. Firstly, the general SINS error model is given and rewrite it as the form of linear state space model. Then, the state vector decomposing and the updating equation of decomposed covariance matrix is deduced in detail. Finally, simulations in the static condition and single axis rotation condition were completed to analyze the contribution of each error sources include initial navigation errors, IMU constant errors and IMU stochastic errors to the final navigation errors.

\section{Error Decomposition Method in System Updating Process}
\subsection{Deterministic System Error distribution}
A deterministic input linear system space state model is
\begin{equation} \label{eq:sysEq1}
    \boldsymbol{\dot{X}} = \boldsymbol{FX}+\boldsymbol{GU}
\end{equation}
in Eq.(\ref{eq:sysEq1}), $\boldsymbol{X}$ is $n\times1$ state vector, $ \boldsymbol{F} $ is system matrix, $ \boldsymbol{G} $ is input distribution matrix, and $\boldsymbol{U}$ is $m\times1$ deterministic input vector of system. $\boldsymbol{X}$ is calculated only by initial state $\boldsymbol{X}_0$ and input vector $\boldsymbol{U}$, if $\boldsymbol{X}$ is used to represent system error of INS. The discrete model function is given by
\begin{equation} \label{eq:sysEq2}
    \boldsymbol{X}_{k} = \boldsymbol{\Phi}_{k/k-1} \boldsymbol{X}_{k-1} + \boldsymbol{\Gamma}_{k-1} \boldsymbol{U}_{k-1}
\end{equation}
in where $\boldsymbol{X}_k$ is the state vector at $t_k$, $\boldsymbol{\Phi}_{k/k-1}$ is state transfer matrix from $t_{k-1}$ to $t_k$, $\boldsymbol{\Gamma}_k$ is discrete input matrix, and $\boldsymbol{U}_{k}$ is input vector at $t_k$.

Assuming that the initial vector of system error(state vector) is $\boldsymbol{X}_0$ and the input vector series is $\boldsymbol{U_0, U_1,\cdots, U_n}$, the system error updating process in the period $k=0\rightarrow n$ is given by
\begin{equation} \label{eq:XUdt1}
    \left\{ \begin{aligned}
        &\boldsymbol{X}_1 = \boldsymbol{X}_{1/0} \boldsymbol{X}_0 + \boldsymbol{\Gamma}_0 \boldsymbol{U}_0 \\
        &\boldsymbol{X}_2 = \boldsymbol{X}_{2/1} \boldsymbol{X}_1 + \boldsymbol{\Gamma}_1 \boldsymbol{U}_1 \\
        \vdots \\
        &\boldsymbol{X}_n = \boldsymbol{X}_{n/n-1} \boldsymbol{X}_{n-1} + \boldsymbol{\Gamma}_{n-1} \boldsymbol{U}_{n-1}
    \end{aligned} \right.
\end{equation}

Substituting the first $n-1$ equation in Eq.(\ref{eq:XUdt1}) into the last equation, we have
\begin{equation}
    \begin{aligned}
    \boldsymbol{X}_{n} =&\boldsymbol{\Phi}_{n \backslash n-1} \boldsymbol{X}_{n-1}+\boldsymbol{\Gamma}_{n-1} \boldsymbol{U}_{n-1} \\
    =&\boldsymbol{\Phi}_{n \backslash n-1}\left(\boldsymbol{\Phi}_{n-1 \backslash n-2} \boldsymbol{X}_{n-2}+\boldsymbol{\Gamma}_{n-2} \boldsymbol{U}_{n-2}\right)+ \\
    &\boldsymbol{\Gamma}_{n-1} \boldsymbol{U}_{n-1} \\
    =&\boldsymbol{\Phi}_{n \backslash n-1} \boldsymbol{\Phi}_{n-1 \backslash n-2} \boldsymbol{X}_{n-2}+ \\
    &\boldsymbol{\Phi}_{n \backslash n-1} \boldsymbol{\Gamma}_{n-2} \boldsymbol{U}_{n-2}+ \boldsymbol{\Gamma}_{n-1} \boldsymbol{U}_{n-1} \\
    & \cdots \\
    =&\prod_{i=1}^{n} \boldsymbol{\Phi}_{i \backslash i-1} \boldsymbol{X}_{0}+\sum_{j=0}\left(\prod_{k=j}^{n-2} \boldsymbol{\Phi}_{k+2 \backslash k+1} \boldsymbol{\Gamma}_{k} \boldsymbol{U}_{k}\right) \\
    &+\boldsymbol{\Gamma}_{n-1} \boldsymbol{U}_{n-1}
    \end{aligned}
\end{equation}

It is clear that the state vector $ \boldsymbol{X}_n $ at $ t_n $ is determined by initial state vector $ \boldsymbol{X}_0 $ and the input vector series from $ t_0 $ to $ t_{n-1} $. Hence, the current state can be divided into the part caused by initial state and the part cause by input vector series.

Make state vector $ \boldsymbol{X}_k $ be divided as
\begin{equation}
	\boldsymbol{X}_k = \boldsymbol{\bar{X}}_k + \boldsymbol{\bar{U}}_k \space,\left(k=0,1,2,\cdots,n\right)
\end{equation}
in where $ \boldsymbol{\bar{X}}_k $ is part of state caused by initial state and $ \boldsymbol{\bar{U}}_k $ is part of state caused by input series. The updating equations of them are given by
\begin{equation}
	\boldsymbol{\bar{X}}_k = \boldsymbol{\Phi}_{k/k-1}\boldsymbol{\bar{X}}_{k-1}, \space \left(\boldsymbol{\bar{X}}_0 = \boldsymbol{X}_0, \space k=0,1,2,\cdots, n\right)
\end{equation}
\begin{equation}
	\boldsymbol{\bar{U}}_k = \boldsymbol{\Phi}_{k/k-1}\boldsymbol{\bar{U}}_{k-1} + \boldsymbol{\Gamma}_{k-1}\boldsymbol{U}_{k-1}
\end{equation}

Hence, when $ k=n $ we have
\begin{equation}
	\boldsymbol{\bar{X}}_n = \prod_{i=1}^{n} \boldsymbol{\Phi}_{i/i-1} \boldsymbol{\bar{X}}_0
\end{equation}
\begin{equation}
	\boldsymbol{\bar{U}}_n = \sum_{j=1}^{n-2} \left(\prod_{k=j}^{n-2} \boldsymbol{\Phi}_{k+2/k+1} \boldsymbol{\Gamma}_{k}\boldsymbol{U}_{k} \right) + \boldsymbol{\Gamma}_{n-1}\boldsymbol{U}_{n-1}
\end{equation}

And the state transfer equation in Eq.(\ref{eq:sysEq2}) is written as
\begin{equation} \label{eq:sysEq3}
	\left\{ \begin{aligned}
		& \boldsymbol{\bar{X}}_{k+1} = \boldsymbol{\Phi}_{k+1/k}\boldsymbol{\bar{X}}_{k} \\
		& \boldsymbol{\bar{U}}_{k+1} = \boldsymbol{\Phi}_{k+1/k}\boldsymbol{\bar{U}}_{k} + \boldsymbol{\Gamma}_{k} \boldsymbol{{U}}_k \\
		& \boldsymbol{X}_{k+1} = \boldsymbol{\bar{X}}_{k+1} + \boldsymbol{\bar{U}}_{k+1}
	\end{aligned}\right. \space \space
	\left( \begin{gathered}
		\boldsymbol{\bar{X}}_0 = \boldsymbol{X}_0 \\
		\boldsymbol{\bar{U}}_0 = \boldsymbol{0} \\
		k=0,1,\cdots,n-1
	\end{gathered}\right)
\end{equation} 

Eq.(\ref{eq:sysEq3}) indicates that the updating process of system state can be divided into the initial state updating and input vector updating. Hence, the proportion in current state vector of initial state and input vectors is obtained easily with the updating equation in Eq.(\ref{eq:sysEq3}). However, the proportion of each element of initial vector and input vector still cannot be obtained. The further decomposition of Eq.(\ref{eq:sysEq1}) should be executed.

If decomposing the state vector $ \boldsymbol{X}_{k} $ as a series of vector $ \boldsymbol{X}_{k,(i)} $, $ \boldsymbol{X}_{k} $ will become
\begin{equation} \label{eq:XkDecmp}
	\boldsymbol{\bar{X}}_{k} = \boldsymbol{\bar{X}}_{k,(1)} + \boldsymbol{\bar{X}}_{k,(2)} + \cdots + \boldsymbol{\bar{X}}_{k,(n)}
\end{equation}
in where $ \boldsymbol{X}_{k,(i)} $ represents part of current state caused by the i-th element in initial state vector. When $ k=0 $, we have
\begin{equation}
	\left\{ \begin{aligned}
		&\boldsymbol{\bar{X}}_{0,(1)} = \left[\begin{matrix}  X_{0,(1)} &  \boldsymbol{0}_{n-1\times1} \end{matrix}\right]^{\mathrm{T}} \\
		\vdots	\\
		&\boldsymbol{\bar{X}}_{0,(i)} = \left[\begin{matrix} \boldsymbol{0}_{i-1\times1} & X_{0,(i)} &  \boldsymbol{0}_{n-i\times1} \end{matrix}\right]^{\mathrm{T}} \\
		\vdots  \\
		&\boldsymbol{\bar{X}}_{0,(n)} = \left[\begin{matrix} \boldsymbol{0}_{n-1\times1} & X_{0,(n)}  \end{matrix}\right]^{\mathrm{T}} 
	\end{aligned} \right.
\end{equation}
in where $ X_{k,(i)} $ is the $ i $-th element in $ \boldsymbol{X}_k $.

Similarly, the input vector series $ \boldsymbol{U}_{i} $and corresponding state vector part $ \boldsymbol{\bar{U}}_{i} $ are also decomposed as
\begin{equation}
	\boldsymbol{\bar{U}}_{k} = \boldsymbol{\bar{U}}_{k,(1)} + \boldsymbol{\bar{U}}_{k,(2)} + \cdots + \boldsymbol{\bar{U}}_{k,(m)}
\end{equation}
\begin{equation}
	\boldsymbol{{U}}_{k} = \boldsymbol{{U}}_{k,(1)} + \boldsymbol{{U}}_{k,(2)} + \cdots + \boldsymbol{{U}}_{k,(m)}
\end{equation}

Finally, the state transfer equation that decomposed by each element is given by
\begin{equation} \label{eq:sysEq4}
	\begin{aligned}
		&\left\{ \begin{aligned}
		& \boldsymbol{\bar{X}}_{k+1,(i)} = \boldsymbol{\Phi}_{k+1/k}\boldsymbol{\bar{X}}_{k,(i)} \\
		& \boldsymbol{\bar{U}}_{k+1,(j)} = \boldsymbol{\Phi}_{k+1/k}\boldsymbol{\bar{U}}_{k},(j) + \boldsymbol{\Gamma}_{k} \boldsymbol{\bar{U}}_{k,(j)} \\
		& \boldsymbol{X}_{k+1} = \sum_{i=1}^{n}\boldsymbol{\bar{X}}_{k+1,(i)} + \sum_{j=1}^{m}\boldsymbol{\bar{U}}_{k+1,(j)}
		\end{aligned}\right. 
		\left( \begin{gathered}
			\boldsymbol{\bar{U}}_{0,(j)} = \boldsymbol{0}
		\end{gathered}\right) \\
		&\left( \begin{gathered}
			i=1,2,\cdots,n; j=1,2,\cdots,m;	k=0,1,\cdots,n-1
		\end{gathered}\right)
	\end{aligned}
\end{equation}

The influence from every element in initial state vector and input series will be represented precisely according to $ \boldsymbol{\bar{X}}_{k,(i)} $ and $ \boldsymbol{\bar{U}}_{k,(i)} $, after updating system with Eq.(\ref{eq:sysEq4}).

\subsection{Stochastic System Error distribution}
When processing deterministic linear systems, the error distribution result is worked out after only once system error model updating with Eq.(\ref{eq:sysEq4}). However, the Monte-Carlo method, that needs statistics navigation errors after enough much simulation, is generally used in error distribution analysis of stochastic system.

A stochastic input linear system space state model is given by
\begin{equation}
	\boldsymbol{\dot{X}} = \boldsymbol{FX} + \boldsymbol{GW}
\end{equation}
in where $ \boldsymbol{W} $ is a stochastic input vector. Similarly, define $ \boldsymbol{X} $ is $ n\times1 $ state vector, and $ \boldsymbol{W} $ is $ m\times1 $ noise vector. Its discrete form is 
\begin{equation} \label{eq:StoSys1}
	\boldsymbol{X}_{k} = \boldsymbol{\Phi}_{k/k-1} \boldsymbol{X}_{k-1} + \boldsymbol{\Gamma}_{k-1} \boldsymbol{W}_{k-1}
\end{equation}

Transpose both sides of the discrete equation, we have
\begin{equation} \label{eq:StoSys1T}
	\boldsymbol{X}_{k}^{\mathrm{T}} = \boldsymbol{X}_{k-1}^{\mathrm{T}} \boldsymbol{\Phi}_{k/k-1}^{\mathrm{T}} + \boldsymbol{W}_{k-1}^{\mathrm{T}} \boldsymbol{\Gamma}_{k-1}^{\mathrm{T}}
\end{equation}

The equation that product Eq.(\ref{eq:StoSys1}) and (\ref{eq:StoSys1T}) is given by
\begin{equation} \label{eq:StoSysProd}
	\begin{aligned}
		\boldsymbol{X}_{k}\boldsymbol{X}_{k}^{\mathrm{T}} =& \left(\boldsymbol{\Phi}_{k/k-1} \boldsymbol{X}_{k-1} + \boldsymbol{\Gamma}_{k-1} \boldsymbol{W}_{k-1}\right)\cdot \\
		&\left(\boldsymbol{X}_{k-1}^{\mathrm{T}} \boldsymbol{\Phi}_{k/k-1}^{\mathrm{T}}  + \boldsymbol{W}_{k-1}^{\mathrm{T}} \boldsymbol{\Gamma}_{k-1}^{\mathrm{T}} \right) \\
		=& \boldsymbol{\Phi}_{k/k-1} \boldsymbol{X}_{k-1} \boldsymbol{X}_{k-1}^{\mathrm{T}} \boldsymbol{\Phi}_{k/k-1}^{\mathrm{T}} + \\
		& \boldsymbol{\Phi}_{k/k-1} \boldsymbol{X}_{k-1} \boldsymbol{W}_{k-1}^{\mathrm{T}} \boldsymbol{\Gamma}_{k-1}^{\mathrm{T}} + \\
		& \boldsymbol{\Gamma}_{k-1} \boldsymbol{W}_{k-1} \boldsymbol{X}_{k-1}^{\mathrm{T}} \boldsymbol{\Phi}_{k/k-1}^{\mathrm{T}} + \\
		 & \boldsymbol{\Gamma}_{k-1} \boldsymbol{W}_{k-1} \boldsymbol{W}_{k-1}^{\mathrm{T}} \boldsymbol{\Gamma}_{k-1}^{\mathrm{T}}
	\end{aligned}
\end{equation}

The expected number of Eq.(\ref{eq:StoSysProd}) is
\begin{equation}
	\begin{aligned}
		\mathrm{E}\left(\boldsymbol{X}_{k}\boldsymbol{X}_{k}^{\mathrm{T}}\right) =& \boldsymbol{\Phi}_{k,k-1}\mathrm{E}\left(\boldsymbol{X}_{k-1}\boldsymbol{X}_{k-1}^{\mathrm{T}}\right) \boldsymbol{\Phi}_{k,k-1}^{\mathrm{T}} + \\ 
		& \boldsymbol{\Gamma}_{k-1} \mathrm{E}\left(\boldsymbol{w}_{k-1}\boldsymbol{w}_{k-1}^{\mathrm{T}}\right) \boldsymbol{\Gamma}_{k-1}^{\mathrm{T}} +\\
		& \boldsymbol{\Gamma}_{k-1} \mathrm{E}\left(\boldsymbol{w}_{k-1}\boldsymbol{X}_{k-1}^{\mathrm{T}}\right) \boldsymbol{\Phi}_{k-1}^{\mathrm{T}} + \\
		&\boldsymbol{\Phi}_{k-1} \mathrm{E}\left(\boldsymbol{X}_{k-1}\boldsymbol{w}_{k-1}^{\mathrm{T}}\right) \boldsymbol{\Gamma}_{k-1}^{\mathrm{T}} \\
		=& \boldsymbol{\Phi}_{k/k-1}\boldsymbol{P}_{k-1}\boldsymbol{\Phi}_{k/k-1}^{\mathrm{T}} + \boldsymbol{\Gamma}_{k-1}\boldsymbol{Q}_{k-1}\boldsymbol{\Gamma}_{k-1}^{\mathrm{T}}
	\end{aligned}
\end{equation}
in where $\boldsymbol{P}_{k}$ is system error covariance matrix, $\boldsymbol{Q}_{k}$is variance matrix of stochastic error.

At the $ n $-th updating time, the system covariance matrix, that has been updated $ n $ times, is expanded as
\begin{equation} \label{eq:Pn}
	\begin{aligned}
		\boldsymbol{P}_{n} =& \boldsymbol{\Phi}_{n/n-1} \boldsymbol{P}_{n-1} \boldsymbol{\Phi}_{n/n-1}^{\mathrm{T}} + \boldsymbol{\Gamma}_{n-1} \boldsymbol{Q}_{n-1} \boldsymbol{\Gamma}_{n-1} \\
		=& \prod_{i=1}^{n}\boldsymbol{\Phi}_{i/i-1}\boldsymbol{P}_{0}\boldsymbol{\Phi}_{i/i-1}^{\mathrm{T}} + \\
		& \sum_{j=0}^{n-2} \left( \prod_{k=j}^{n-2} \boldsymbol{\Phi}_{k+2/k+1} \boldsymbol{\Gamma}_{k} \boldsymbol{Q}_{k} \boldsymbol{\Gamma}_{k}^{\mathrm{T}} \boldsymbol{\Phi}_{k+2/k+1}^{\mathrm{T}} \right)+ \\
		& \boldsymbol{\Gamma}_{n-1} \boldsymbol{Q}_{n-1} \boldsymbol{\Gamma}_{n-1}^{\mathrm{T}}
	\end{aligned}
\end{equation}

According to Eq.(\ref{eq:Pn}), the $ n $-th covariance matrix result is only related with initial covariance matrix and the stochastic input series. Hence, if define
\begin{subequations}
	\begin{equation}
		\begin{gathered}
			\bar{\boldsymbol{P}}_{k+1} = \boldsymbol{\Phi}_{k+1/k} \bar{\boldsymbol{P}}_{k} \boldsymbol{\Phi}_{k+1/k}^{\mathrm{T}} , \\
			(\bar{\boldsymbol{P}_{0}} = \boldsymbol{P}_{0})(k=0,1,2,\cdots,n-1)
		\end{gathered}
	\end{equation}
	\begin{equation}
		\begin{gathered}
			\bar{\boldsymbol{Q}}_{k+1} = \boldsymbol{\Phi}_{k+1/k} \bar{\boldsymbol{Q}}_{k} \boldsymbol{\Phi}_{k+1/k}^{\mathrm{T}} + \boldsymbol{\Gamma}_{k} {\boldsymbol{Q}}_{k} \boldsymbol{\Gamma}_{k}^{\mathrm{T}} ,\\
			(\bar{\boldsymbol{Q}_{0}} = \boldsymbol{0})(k=0,1,2,\cdots,n-1)
		\end{gathered}
	\end{equation}
\end{subequations}

Eq.{\ref{eq:Pn}} will be rewritten as below.
\begin{equation}
	\bar{\boldsymbol{P}}_{n} = \prod_{i=1}^{n} \left(\boldsymbol{\Phi}_{i/i-1} \boldsymbol{P}_{0} \boldsymbol{\Phi}_{i/i-1}^{\mathrm{T}}\right)
\end{equation}
\begin{equation}
	\begin{aligned}
		\bar{\boldsymbol{Q}}_{n} =& \sum_{j=0}^{n-2} \left( \sum_{k=j}^{n-2} \boldsymbol{\Phi}_{k+1/k+1} \boldsymbol{\Gamma}_{k} \boldsymbol{Q}_{k} \boldsymbol{\Gamma}_{k}^{\mathrm{T}} \boldsymbol{\Phi}_{k+1/k+1}^{\mathrm{T}} \right) +\\ &\boldsymbol{\Gamma}_{n-1} \boldsymbol{Q}_{n-1} \boldsymbol{\Gamma}_{n-1}^{\mathrm{T}}
	\end{aligned}
\end{equation}

The above results can be summarized as follows covariance matrix updating equation in Eq.(\ref{eq:PkSys}).
\begin{equation} \label{eq:PkSys}
	\begin{gathered}
	\left\{ \begin{aligned}
		&\bar{\boldsymbol{P}}_{k+1} = \boldsymbol{\Phi}_{k+1/k} \bar{\boldsymbol{P}}_{k} \boldsymbol{\Phi}_{k+1/k}^{\mathrm{T}} \\
		&\bar{\boldsymbol{Q}}_{k+1} = \boldsymbol{\Phi}_{k/k-1} \bar{\boldsymbol{Q}}_{k} \boldsymbol{\Phi}_{k/k-1}^{\mathrm{T}} + \boldsymbol{\Gamma}_{k} \boldsymbol{Q}_{k} \boldsymbol{\Gamma}_{k}^{\mathrm{T}} \\
		&\boldsymbol{P}_{k+1} = \bar{\boldsymbol{P}}_{k+1}+\bar{\boldsymbol{Q}}_{k+1}
	\end{aligned}\right. \space,
    \left( \begin{gathered}
    	\bar{\boldsymbol{P}}_{0} = \boldsymbol{P}_{0} \\
    	\bar{\boldsymbol{Q}}_{0} = \boldsymbol{0}     	
    \end{gathered} \right) \\
	\left(k=0,1,\cdots,n-1\right)
	\end{gathered}
\end{equation}

Similar with Eq.(\ref{eq:XkDecmp}), make Eq.(\ref{eq:PkSys}) be decomposed according to each element on diagonal of covariance matrix as Eq.(\ref{eq:PkDecmp}).
\begin{subequations} 
	\label{eq:PkDecmp}
	\begin{equation}
	\bar{\boldsymbol{P}}_{k} = \bar{\boldsymbol{P}}_{k,(1)} +  \bar{\boldsymbol{P}}_{k,(2)} + \cdots + \bar{\boldsymbol{P}}_{k,(p)}
	\end{equation}
	\begin{equation}
	\bar{\boldsymbol{Q}}_{k} = \bar{\boldsymbol{Q}}_{k,(1)} +  \bar{\boldsymbol{Q}}_{k,(2)} + \cdots + \bar{\boldsymbol{Q}}_{k,(m)}
	\end{equation}
\end{subequations}

Covariance matrix updating equation (\ref{eq:PkSys}) is decomposed as:
\begin{equation}
	\begin{gathered}
	\left\{ \begin{aligned}
		&\bar{\boldsymbol{P}}_{k+1,(i)} = \boldsymbol{\Phi}_{k+1/k} \bar{\boldsymbol{P}}_{k,(i)} \boldsymbol{\Phi}_{k+1/k}^{\mathrm{T}} \\
		&\bar{\boldsymbol{Q}}_{k+1,(j)} = \boldsymbol{\Phi}_{k/k-1} \bar{\boldsymbol{Q}}_{k,(j)} \boldsymbol{\Phi}_{k/k-1}^{\mathrm{T}} + \boldsymbol{\Gamma}_{k} \boldsymbol{Q}_{k,(j)} \boldsymbol{\Gamma}_{k}^{\mathrm{T}} \\
		&\boldsymbol{P}_{k+1} = \sum_{i=1}^{p}\bar{\boldsymbol{P}}_{k+1,(j)} + \sum_{j=1}^{m}\bar{\boldsymbol{Q}}_{k+1,(j)}
	\end{aligned}\right. \\
	(i=1,\cdots,p; j=1,\cdots,m; k=0,1,\cdots, n-1)
	\end{gathered}
\end{equation}

After the whole update process, a group of covariance matrix $ \bar{\boldsymbol{P}}_{k,(1)}, \cdots, \bar{\boldsymbol{P}}_{k,(p)}, \bar{\boldsymbol{Q}}_{k,(1)}, \cdots, \bar{\boldsymbol{Q}}_{k,(m)} $ represents part caused by each initial variance element and stochastic error is obtained.

\section{Algorithm Evaluation}

The stochastic error distribution method will be applied to analysis error transformation of SINS error model, because the errors here must be described as variance matrix to analysis the effect from noise of gyroscope and accelerometer.

The initial matrix of $ \boldsymbol{P}_{0} $ is a diagonal matrix, and diagonal elements are given by
\begin{equation}
	\boldsymbol{P}_{0}(i,i) = \boldsymbol{X}_{0}(i,i)^2
\end{equation}
in where $ \cdot(i,i) $ represents the element in the $ i $-th row and $ i $-th column.

According to the SINS error model, $ \boldsymbol{P}_{k} $ can be decomposed with 13 covariance  matrix of initial errors and 6 covariance matrix of stochastic errors. Hence, we have
\begin{equation}
	\boldsymbol{P}_{k} = \sum_{i=1}^{28} \bar{\boldsymbol{P}}_{k,(i)} + \sum_{j=1}^{6} \bar{\boldsymbol{Q}}_{k,(j)}
\end{equation}

The correspondence between error source and serial number $ i $ and $ j $ are listed in Table \ref{tab:error}.

\begin{table}[!ht]
	\centering
	\caption{Correspondence of error source and serial number}
	\label{tab:error}
	\begin{tabular}{m{1cm}m{3cm}m{3cm}}
		\hline
		Num. & Error source & signature \\
		\hline
		1,2,3 & attitude error & $ \phi_{E},\phi_{N},\phi_{U} $ \\
		4,5 & velocity error & $ \delta v_{E},\delta v_{N},\delta v_{U} $ \\
		6,7 & position error & $ \delta L,\delta \lambda $ \\
		\hline
		8,9,10 & gyroscope scale factor error & $ \delta K_{g,11},\delta K_{g,22},\delta K_{g,33} $ \\
		11,12,13 & gyroscope mounting error & $ \delta K_{g,21},\delta K_{g,31},\delta K_{g,32} $ \\
		14,15,16 & accelerometer scale factor error & $ \delta K_{g,11},\delta K_{g,22},\delta K_{g,33} $ \\
		17\textasciitilde22 & accelerometer mounting error & $ \begin{aligned}
			\delta K_{g,12},\delta K_{g,13},\delta K_{g,21},\\ \delta K_{g,23},\delta K_{g,31},\delta K_{g,32}
		\end{aligned} $ \\
		23,24,25 & gyroscope bias & $ \varepsilon_{x},\varepsilon_{y},\varepsilon_{z} $ \\
		26,27,28 & accelerometer bias & $ \nabla_{x},\nabla_{y},\nabla_{z} $ \\
		\hline
		1,2,3 & gyroscope & $ w_{g,x},w_{g,y},w_{g,z} $ \\
		4,5,6 & accelerometer & $ w_{a,x},w_{a,y},w_{a,z} $ \\
		\hline
	\end{tabular}
\end{table}

The proposed error distribution method can easily analyze the error propagation law according to the finial state covariance matrices $ \bar{\boldsymbol{P}}_{k,(i)} $ and $ \bar{\boldsymbol{Q}}_{k,(j)} $. This method is much different from conventional Monte-Carlo method, because the new method doesn't need lots of simulation and obtaining stochastic error proportion from final statistical result. This is a simple, fast and feasible error distribution method for SINS error analysis.

In the next algorithm evaluation, the error distribution result will be obtained with the proposed method as bar charts of navigation error proportion.

\subsection{Test of static navigation simulation}
In static condition, most parameters of system matrix $ \boldsymbol{F} $ are constant values. The SINS error model will degenerate into a simple form,that the effect of each error to navigation error will be obtained directly. The simulation result of proposed method can be evaluated by comparing with the analysis based on simple form of SINS error model.

\begin{figure*}[!ht]
	\normalsize
	\centering
	\subfigure[Error distribution result of attitude error]{ {\includegraphics[width=13cm]{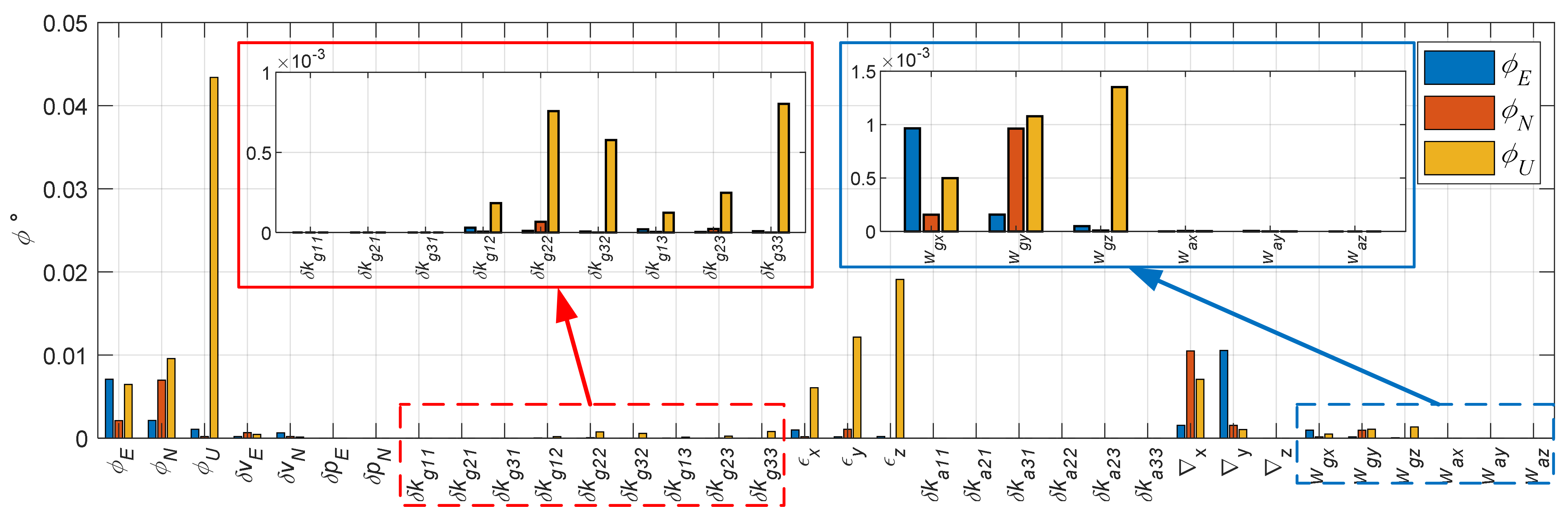}}}
	\subfigure[Error distribution result of velocity error]{ {\includegraphics[width=13cm]{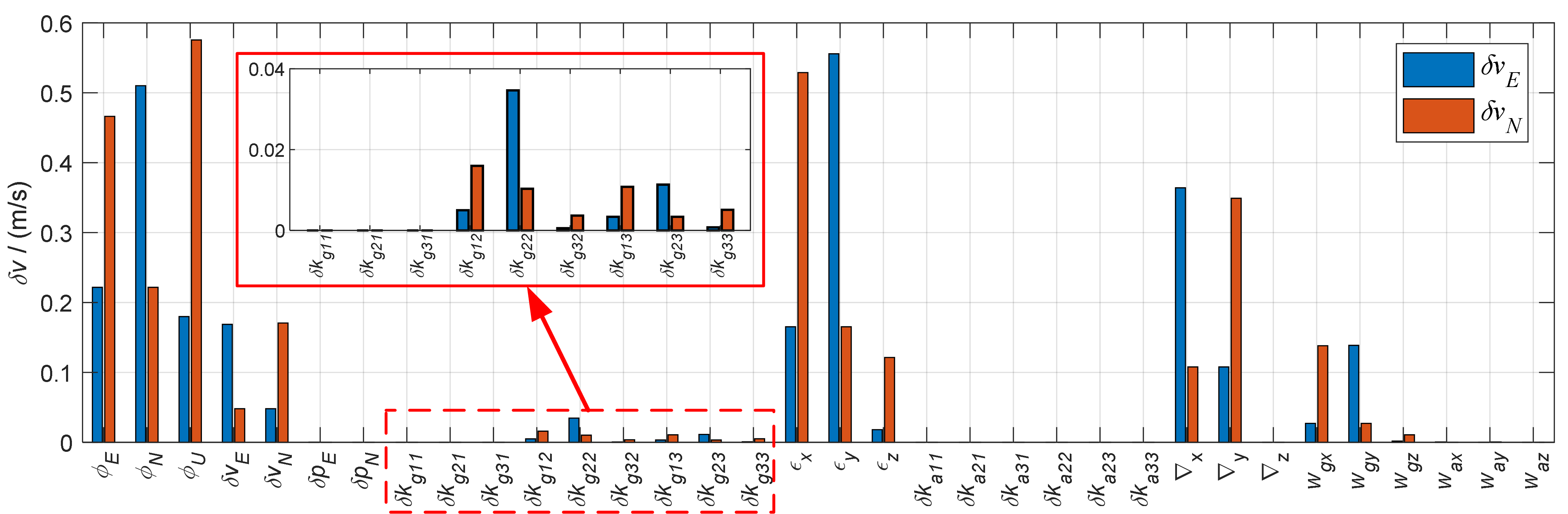}}}	
	\subfigure[Error distribution result of position error]{ {\includegraphics[width=13cm]{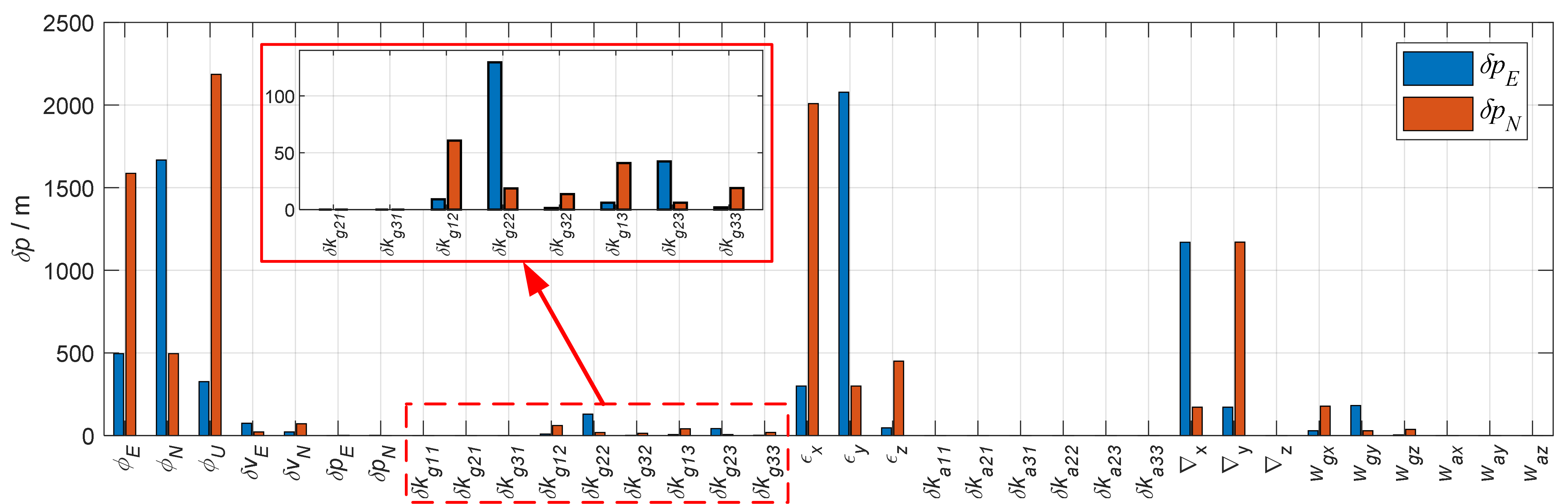}}}
	\caption{Error distribution result of simulation in static condition}
	\label{fig:sim1}
\end{figure*}

Part of the specification of SIMU(Strapdown Inertial Measurement Unit) in the simulation is shown in Table \ref{tab:specif_IMU}. And the distribution results of attitude errors , velocity errors and position errors are shown in Fig.\ref{fig:sim1}.
\begin{table}[!th]
	\centering
	\caption{The specification of SIMU in the simulation}
	\label{tab:specif_IMU}
	\begin{tabular}{m{5cm}m{3cm}}
		\hline
		Parameter & Values \\
		\hline
		Frequency of sampling & 100Hz \\
		Initial error of attitude & $ 30^{\prime\prime},30^{\prime\prime},3^{\prime}$ \\
		Initial error of velocity & 0.2m/s,0.2m/s \\
		Initial error of position & 2m,2m \\
		Bias of gyroscope & $ 0.01^\circ\mathrm{/h} $ \\
		Bias of accelerometer & $100\mathrm{\mu g} $ \\
		Scale factor error of gyroscopes & 50ppm \\
		Scale factor error of accelerometers & 50ppm \\
		Mounting error of gyroscopes &  $ 5^{\prime\prime} $ \\
		Mounting error of accelerometers & $ 5^{\prime\prime} $ \\
		Random walk of angle & $ 0.001^\circ/\sqrt{\mathrm{h}} $\\
		Random walk of velocity & $1\mathrm{\mu g}/\sqrt{\mathrm{Hz}} $\\
		\hline
	\end{tabular}
\end{table}

The conclusions below could be indicted from the simulation:

(1) Initial attitude error, initial velocity error and the bias errors of gyroscopes and accelerometers are primary error source. In the static condition, scale factor error and mounting error make less effect to final navigation error.

(2) The static attitude error model of SINS is given by
\begin{equation}
	\label{eq:dPhi}
	\dot{\boldsymbol{\phi}} = \boldsymbol{\phi} \times \boldsymbol{\omega}_{ie}^{n} - \boldsymbol{\varepsilon}^n
\end{equation}
In Eq.(\ref{eq:dPhi}), the attitude error $ \boldsymbol{\phi} $ is mainly produced by the initial attitude error and bias error. And this characteristic can be reflected from the attitude error distribution result.

(3) The static velocity error model of SINS is given by
\begin{equation}
	\label{eq:dvn}
	\delta\boldsymbol{\dot{v}}^n = \boldsymbol{g}^n \times \boldsymbol{\phi} + \boldsymbol{C}_b^n \delta\boldsymbol{K}_\mathrm{g} \boldsymbol{C}_n^b \boldsymbol{g}^n + \boldsymbol{\nabla}^n
\end{equation}
In Eq.(\ref{eq:dvn}), attitude error makes gravity incorrectly project on horizontal velocities, so the initial attitude error, the bias of gyroscopes are the primary sensor error. Meanwhile, the bias of accelerometers effect horizontal velocity errors directly, and they also have high percent in velocity error. What indicted from velocity error distribution result is the same as the analysis from Eq.(\ref{eq:dvn}).

(4) The static position error model of SINS is given by
\begin{equation}
	\label{eq:dpos}
	\left\{ \begin{aligned}
		&\delta\dot{L} = \frac{1}{R_M+h}\delta v_N \\
		&\delta\dot{\lambda} = \frac{\sec L\delta v_E}{R_N+h}
	\end{aligned}\right.
\end{equation}
It is clear that position error distribution result will be much similar with velocity error. The velocity error distribution result and position error distribution result have same error proportion distribution in Fig.\ref{fig:sim1}.

\subsection{Test of single axis rotation simulation}
\begin{figure}[!ht]
	\centering
	\includegraphics[width=\columnwidth]{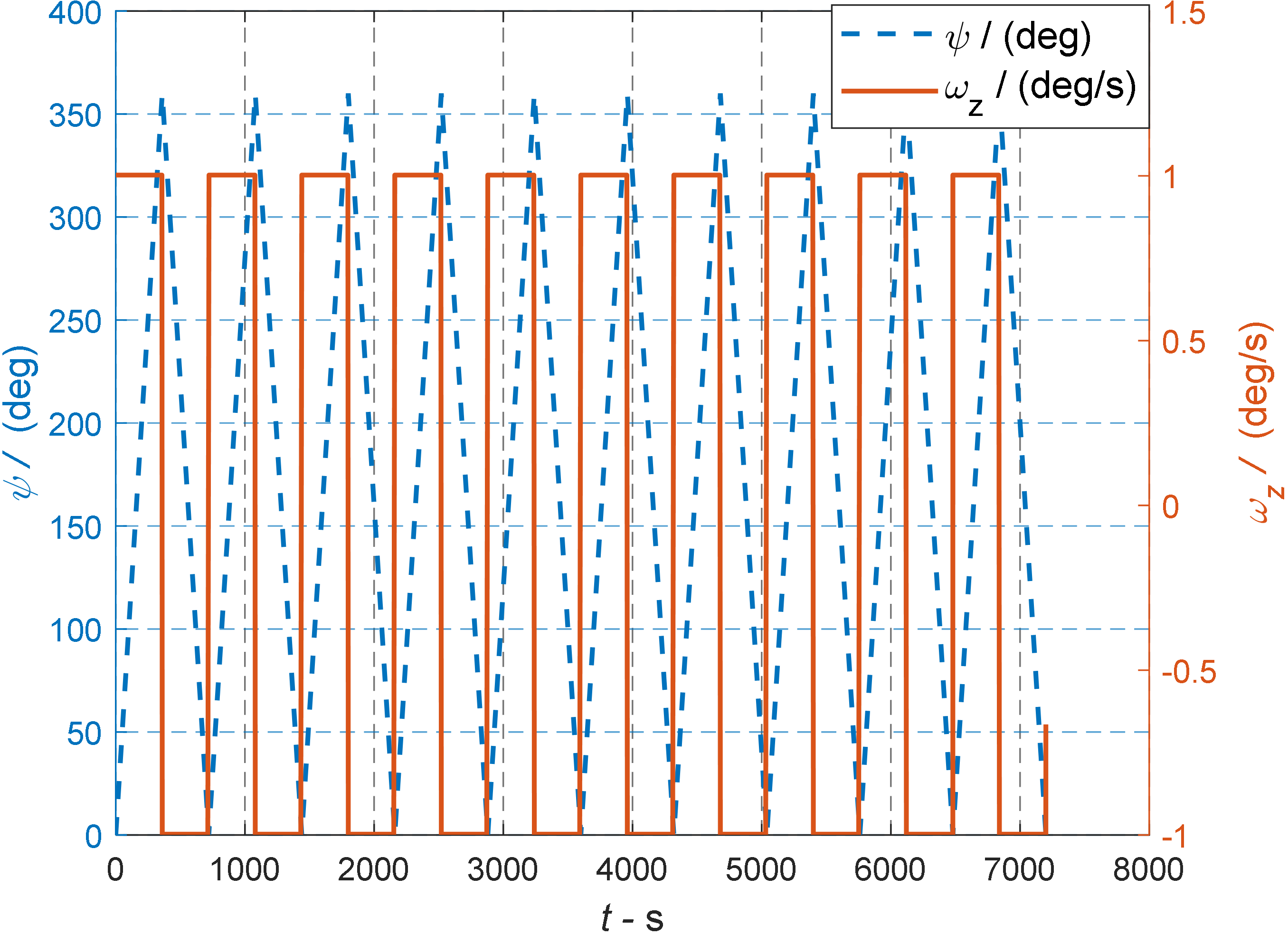}
	\caption{Angular rate and yaw in single axis rotation simulation}
	\label{fig:rotation}
\end{figure}

To compare with the navigation error distribution result of the static test, another error distribution test was finished in the single axis rotation condition with repeated clockwise and anti-clockwise turn. In this motion, the angular rate and yaw curve is in Fig.\ref{fig:rotation}. Rotating about z-axis can restrain the navigation error caused by gyroscope bias and accelerometer bias\cite{Du2016}. Comparing the error distribution result and the theoretical performance restraining navigation error of rotation INS could be used to verify the proposed analysis method.

\begin{figure*}[!ht]
	\normalsize
	\centering
	\subfigure[Error distribution result of attitude error]{ {\includegraphics[width=13cm]{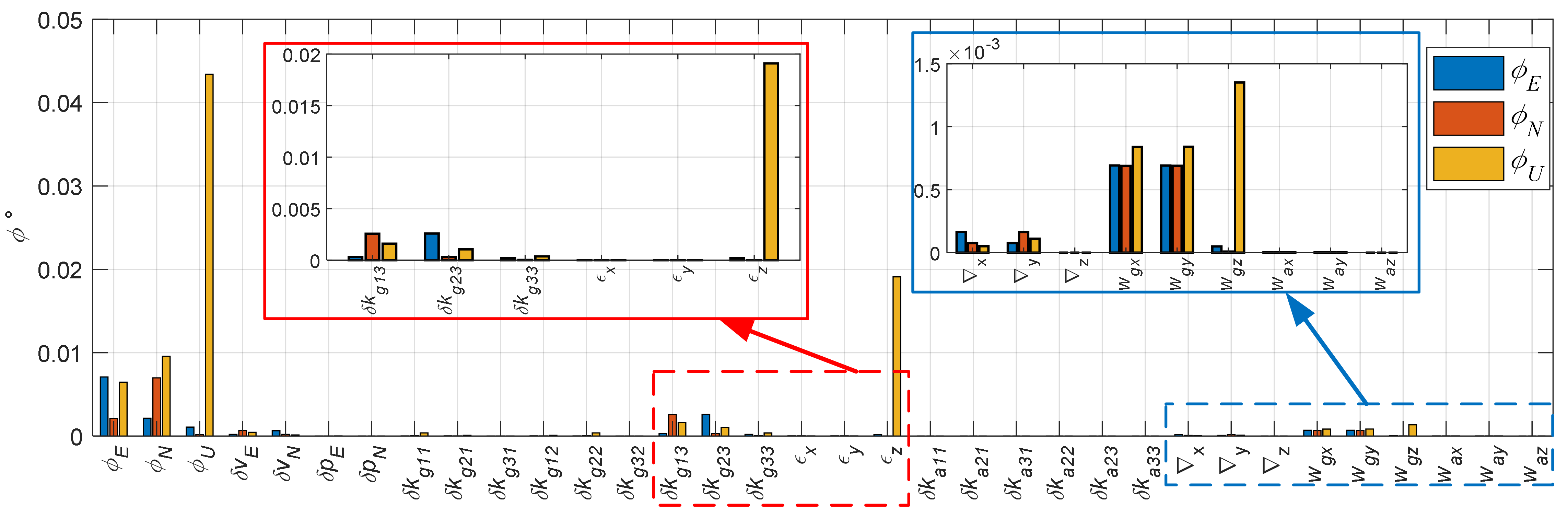}}}
	\subfigure[Error distribution result of velocity error]{ {\includegraphics[width=13cm]{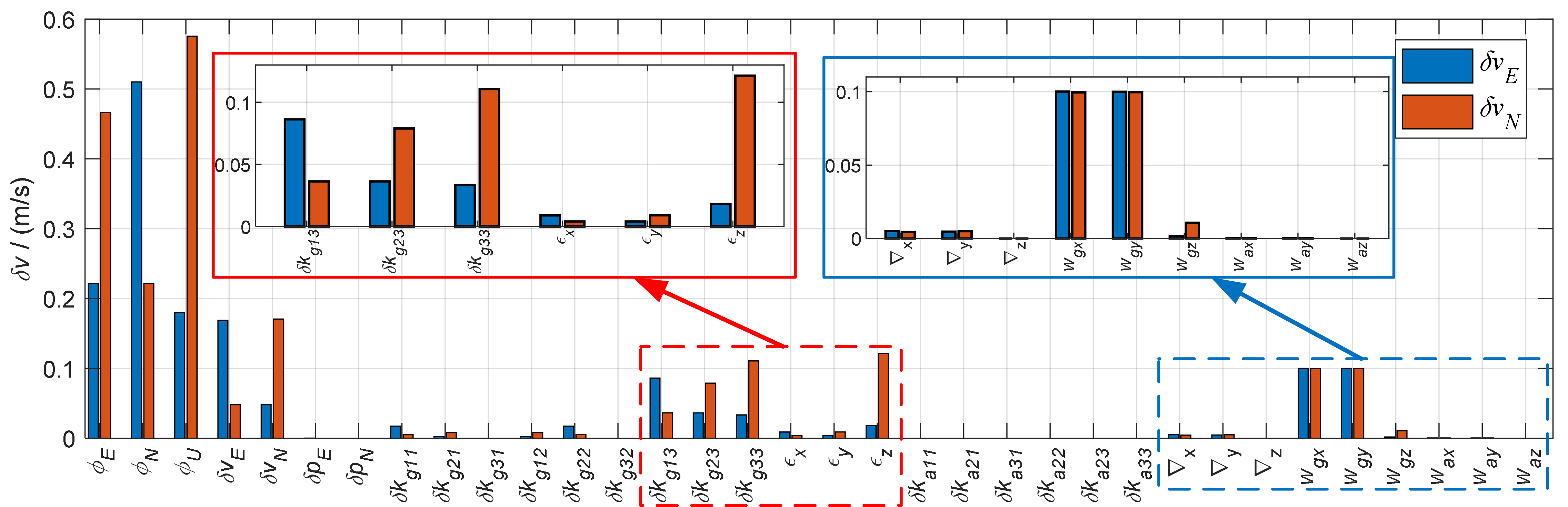}}}	
	\subfigure[Error distribution result of position error]{ {\includegraphics[width=13cm]{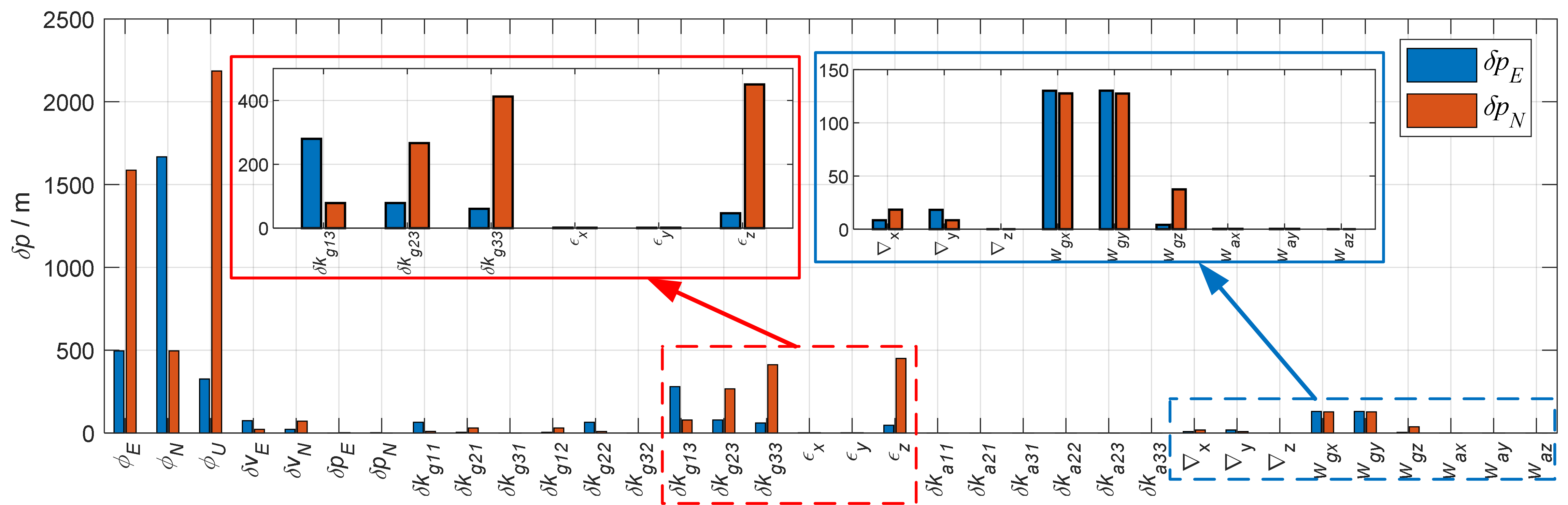}}}
	\caption{Error distribution result of simulation in single axis rotation condition}
	\label{fig:sim2}
\end{figure*}

The conclusions below could be indicted from the simulation:

(1) According to the result of Fig.\ref{fig:sim2}, the influence of horizontal gyroscope bias and accelerometer bias to horizontal attitude error is restrained observably. But the yaw error is almost the same as the static condition. It is in accordance with the regulation of the rotation INS about z-axis.

(2) In the error distribution result of velocity error, horizontal gyroscope bias and accelerometer bias is restrained, however the velocity error caused by scale factor error and mounting error of z-axis is enhanced, because the additional rotation about z-axis 
excites the scale factor error and mounting error.

(3) The error distribution result of rotation INS simulation is correct and in accordance with the error transfer rule of rotation INS. The proposed error distribution analysis method is useful to assist design and optimize the INS.

\section{Conclusion}
After a state space model constructed according to SINS error model, the state vector decomposition of deterministic error is extended to the covariance matrix decomposition with random error, and the decomposed covariance contains error distribution information. Thus, the error distribution of SINS is calculated by updating decomposed covariance matrix with typical trajectory data.

The method of error distribution analysis based on covariance decomposition presented herein can be applied to the design or  pre-research stage of SINS. An error distribution table for multiple error sources can be obtained without any practical test. In the following research stage, the accuracy of SINS can be improved quickly and effectively by optimizing the primary error sources. Therefore, the proposed method has great practical meaning in engineering.

\bibliography{ref.bib}

\end{document}